\documentstyle[amssymb,preprint,aps,floats,epsf,prb,tighten]{revtex}

\newcommand{\ie}{\mbox{i.\ e.\ }}
\newcommand{\eg}{\mbox{e.\ g.\ }}

\newcommand{\etc}{\mbox{\it etc.\ }}
\newcommand{\be}{\begin{eqnarray}}
\newcommand{\en}{\end{eqnarray}}
\newcommand{\no}{\nonumber}
\newcommand{\hc}{\mbox{\it h.\ c.\ }}
\newcommand{\mc}{\mathcal}
\newcommand{\f}{f}
\newcommand{\p}{p}
\newcommand{\fd}{f^{\dagger}}
\newcommand{\pd}{p^{\dagger}}
\newcommand{\at}{\tilde{a}}

\newcommand{\atd}{\tilde{a}^{\dagger}}
\newcommand{\cd}{c^{\dagger}}

\begin{document}
\draft

\title{
Quantum Spin dynamics of
the Bilayer Ferromagnet
La$_{1.2}$Sr$_{1.8}$Mn$_2$O$_7$
}
\author{Nic Shannon}
\address{
Max--Planck--Institut f{\"u}r Physik komplexer Systeme,
N{\"o}thnitzer Str. 38, 01187 Dresden, Germany.
}
\author{Tapan Chatterji}
\address{
Institut Laue-Langevin, BP 156, 38042 Grenoble Cedex 9, France
}
\author{Fatiha Ouchni}
\address{
Institut f{\"u}r Theoretische Physik III,
Universit{\"a}t Stuttgart,
Pfaffenwaldring 57, D-70550 Stuttgart, Germany.
}
\author{Peter Thalmeier}
\address{
Max-Planck-Institut f\"ur chemische Physik fester Stoffe,
N{\"o}thnitzer Str. 40, 01187 Dresden, Germany.
}

\date{\today}
\maketitle
\begin{abstract}
We construct a theory of spin wave excitations
in the bilayer manganite La$_{1.2}$Sr$_{1.8}$Mn$_2$O$_7$
based on the simplest possible double--exchange model,
but including leading quantum corrections to the spin wave
dispersion and damping.
Comparison is made with recent inelastic neutron scattering experiments.
We find that quantum effects account for some part of the
measured damping of spin waves, but cannot by themselves explain
the observed softening of spin waves at the zone boundary. 
Furthermore a doping dependence of the total spin wave dispersion 
and the optical spin wave gap is predicted.
\end{abstract}

\vspace{0.5cm}

\newpage

\section{Introduction}
\label{introduction}

The colossal magnetoresistance (CMR) manganites, of which
perhaps the best known is La$_{1-x}$Ca$_x$MnO$_3$, have been
challenging the theoretical understanding of the way in which
magnetism and metallic behaviour co--exist for more than fifty years.
These materials are difficult to
describe for precisely the very same reason that they are
interesting; namely that they exhibit a complex interplay between
lattice, charge, orbital and spin degrees of freedom.  This gives
rise to a very rich phase diagram, exhibiting different magnetic,
orbital and charge orders, and both metallic and insulating behaviour
as a function of temperature, pressure, applied field and doping
\cite{Rao98,Izyumov01,Chatterji02}.
Even within the ``simple'' low temperature ferromagnetic phase, the
mechanism for the metal--insulator transition which occurs as
ferromagnetic order breaks down remains controversial.

The CMR manganites share a layered perovskite structure with the
even more widely studied high- temperature (HTc) superconductors;
they may be synthesised with one, two,
three (or many), neighbouring conducting planes.  The materials
most frequently discussed are the three dimensional ``infinite layer'' 
compounds, which have equally spaced planes and are approximately cubic
in symmetry.
Here we will construct a theory for ferromagnetism
in La$_{2-2x}$Sr$_{1+2x}$Mn$_2$O$_7$ and discuss results especially for
x=0.4.  In this bilayer compound planes of magnetic
Mn atoms in MnO$_6$ octahedra are grouped in well separated pairs.
The small spin wave dispersion found empirically 
perpendicular to these planes provides us with a justification for considering,
as a first approximation, only a single pair of planes \ie a single
ferromagnetic bilayer with moments lying in the ab--
plane\cite{Hirota98,Chatterji02}. The T--x phase diagram and evolution
of magnetic structure with doping has been reported in
\cite{Kubota00,Okamoto01,Dho01,Hirota01} and a FM phase persists in the range
0.3$\leq x\leq $0.4. For larger doping an intra-- bilayer canting of
moments appears and the charge ordered stoichiometric compound (x=0.5)
LaSr$_2$Mn$_2$O$_7$ finally is an AF insulator. Here we will concentrate
on predictions for the spin wave dispersion and damping of FM bilayer
manganites which have been measured by inelastic neutron scattering
\cite{Hirota01,Chatterji99a,Chatterji99b,Chatterji01a,Fujioka99,Chaboussant00,Perring01}.
Calculation of the spin wave damping requires
going beyond the usual semi--classical picture used to describe
spin wave excitations in the manganites to include quantum effects.
In Section \ref{model} we present a minimal model of a bilayer manganite
based on Zener's double exchange (DE) mechanism \cite{Zener51,Furukawa96}.
A fully quantum mechanical large $S$ expansion of this model is
developed, following a recently introduced operator expansion method
\cite{Shannon01a,Shannon01b}.
Predictions for the dispersion of the optical and acoustic
spin wave modes of a double exchange bilayer, their doping dependence
together with their damping, are made in Section \ref{theory}. A comparison
with
experimental data for La$_{1.2}$Sr$_{1.8}$Mn$_2$O$_7$ is made in Section
\ref{experiment}. This comparison provides a test of how well the
DE model describes FM in CMR materials when quantum effects are included.
We conclude in Section \ref{conclusions} with
a discussion of the implications of our results for the
theory of ferromagnetism in CMR manganites.

\section{The Model Hamiltonian}
\label{model}

In this section we consider La$_{1.2}$Sr$_{1.8}$Mn$_2$O$_7$,
as a concrete example of a bilayer DE system, and derive a model
Hamiltonian for a single La$_{1.2}$Sr$_{1.8}$Mn$_2$O$_7$ bilayer starting
from Zener's DE mechanism, in the limit where the strength of the
Hund's rule coupling is taken to be infinite.  The comparison of the
predictions of this model with experimental data in
Section \ref{experiment} therefore provide a test of how well the
DE model describes FM in CMR materials.
The crystal structure of La$_{2-2x}$Sr$_{1+2x}$Mn$_2$O$_7$ belongs to
the space group I4/mmm with a body centred tetragonal conventional
unit cell that contains two distorted MnO$_6$ octahedra as basis whose
distortion depends on doping \cite{Kubota00,Okamoto01}. The lattice
constants are a =3.87 $\AA$ and c= 20.14 $\AA$. The intra-- bilayer
spacing d$\simeq$a is much smaller than the distance D=6.2 $\AA$
between two adjacent bilayers.
Therefore bilayers are well separated, and the spin wave
spectrum measured by inelastic neutron scattering indeed shows a very
small dispersion of about 0.4 meV in the direction perpendicular to the
planes\cite{Chatterji99a,Rosenkranz00}. For this reason we
will neglect coupling between the planes entirely, and model
La$_{1.2}$Sr$_{1.8}$Mn$_2$O$_7$ in terms of a single pair of layers.
Within a given bilayer, both magnetism and metallic behaviour originate
in the Mn $d$--electrons.
Mn $t_{2g}$ $d$--orbitals are exactly half filled, and form a spin
$3/2$ local moment because of strong Hund's rule coupling.
This local moment couples to
itinerant electron $e_g$ $d$--orbitals through a similar
Hund's rule exchange interaction.
In the metallic phases of the manganites,
electrons in $e_g$ orbitals delocalise by hopping  between
Mn atoms through intermediate O$_{2p}$ orbitals --- a
process named ``double exchange'' by Zener \cite{Zener51}.
This delocalisation of the $e_g$ electrons stabilises FM
order among the $t_{2g}$ spins, since both are strongly coupled
by Hund's rule interaction, and the $e_g$ electrons will have
the maximum kinetic energy if all $t_{2g}$ spins are aligned.

In the bilayer compounds the MnO$_6$ octahedra show a doping
dependent pronounced Jahn-Teller (JT) distortion
\cite{Kubota00,Okamoto01}, therefore Mn$^{3+}$ site symmetry is no longer cubic
and a crystalline electric field (CEF) splitting of e$_g$
(d$_{3z^2-r^2}$,d$_{x^2-y^2}$) states ensues. The influence of this
splitting on the stability of magnetic phases was investigated by Okamoto et al
\cite{Okamoto01}. The e$_g$ splitting energy $\Delta$ is generally smaller than
the inter-site in--plane hopping t and therefore in the FM ground state
the orbital state is of uniformly mixed d$_{3z^2-r^2}$/d$_{x^2-y^2}$
character.  In
this case orbital degrees of freedom do not appear explicitly in the
Hamiltonian but the degree of admixture determines the ratio of
interlayer (t$_\perp$) to intra--layer hopping (t) of the effective single band
(orbital) Hamiltonian which is then given by

\be
\label{eqn:DE}
{\mc H}_{DE} &=& - t \sum_{\langle ij\rangle \lambda \alpha}
      c_{i\lambda\alpha}^{\dagger}c_{j\lambda\alpha}
   - t_{\perp} \sum_{i\alpha} \left\{
      c_{i 1\alpha}^{\dagger}c_{i 2\alpha}
      + \hc \right\}
   - \frac{J_{H}}{2} \sum_{i \lambda \alpha\beta}
      \vec{S}_{i\lambda} . \cd_{i\lambda\alpha}
      \vec{\sigma}_{\alpha\beta}
      c_{i\lambda\beta}
\en
where $c_{\lambda i\alpha}^{\dagger}$ is the creation operator for
an $e_{g}$ electron on site $i$ of plane $\lambda = \{1,2\}$ with spin
$\alpha = \{\uparrow,\downarrow\}$. The components of the operator
$\vec{\sigma}_{\alpha\beta}$ are Pauli matrices, and
$\vec{S}_{i\lambda}$ is the spin operator for the $t_{2g}$
electrons on that site. The on-site exchange $J_H$ parameterises
Hund's rule coupling, and the sum
$\langle ij\rangle$ runs over nearest neighbours within a plane.
Our subsequent DE spin wave analysis will lead to t$\simeq$ 0.175 eV and
t$_\perp\simeq$ 0.1 eV. This is much smaller than the intra-atomic (Hund's
rule) exchange J$_H\sim$ 2 eV which may be estimated from the
splitting of majority and minority spin LDA bands in the
stoichiometric (x=0.5) compound\cite{deBoer99}.

\begin{figure}[tb]
\begin{center}
\leavevmode
\epsfysize = 50.0mm
\epsffile{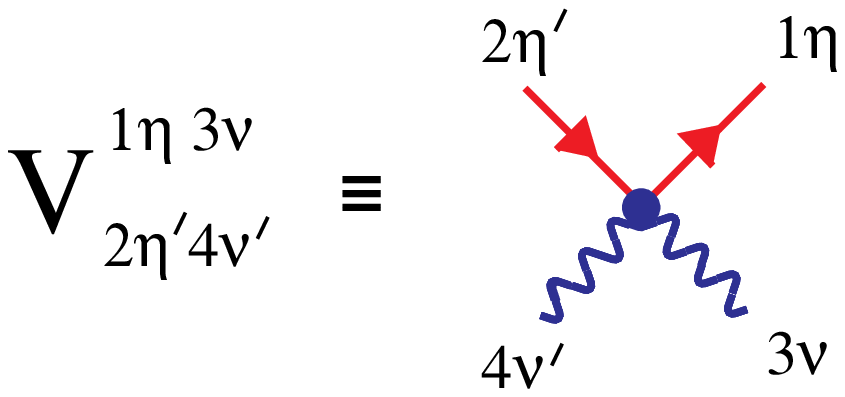}
\caption{
Convention for labelling vertex for interaction between electrons
and spin waves in the limit $J_H/t \to \infty$.
Straight lines correspond to electrons $\f_{k\eta}$
and wavy lines to spin waves $\at_{q\nu}$, where $k$ and $q$ are
momenta in--plane and $\eta,\nu = 0,\pi$ are the momenta out of
plane.
}
\label{fig:vertex}
\end{center}
\end{figure}

In addition, there may be super-exchange interactions between spins,
both within the plane ($J$) and between the two planes of the bilayer
($J_{\perp}$).  These can be parameterised by
\be
\label{eqn:EX}
{\mc H}_{EX} &=& -J^{EX} \sum_{\langle ij\rangle \lambda}
      \vec{T}_{i\lambda } . \vec{T}_{j\lambda}
   - J^{EX}_{\perp} \sum_{i} \left\{
      \vec{T}_{i 1} . \vec{T}_{i 2}
      + \hc \right\}
\en
where
$\vec{T}_{i\lambda }
   = \vec{S}_{i\lambda } + 1/2 \sum_{\alpha\beta}
   c_{i\lambda\alpha}
   \vec{\sigma}_{\alpha\beta}
   c_{i\lambda\beta}
$
is the total spin operator for both $t_{2g}$ and $e_{g}$ $d$--electrons
on the site $i\lambda$.  Exchange integrals in the manganites can be FM
or antiferromagnetic (AF) depending on the details of orbital
occupancy and electronic structure.

To evaluate the spectrum, or even to find the ground state of the
Hamiltonian Eqn.~\ref{eqn:DE} is a formidable task, but if we
assume FM order and
treat the length of the local moment $S$, and the ratio $J_H/t$ as
large parameters, we can derive a controlled expansion of the
properties of a bilayer ferromagnet. The most direct
way of doing this is to work with eigenstates of the Hund's rule
coupling term, and to quantize small fluctuations of the total spin
operator $\vec{T}_{i\lambda }$ using a generalisation
of the usual Holstein--Primakoff procedure due to Shannon and Chubukov
\cite{Shannon01b}.  This approach will now be extended to the
bilayer system.  In the limit  $t/J_H \ll 1$, $t_{\perp}/J_H \ll 1$ $J_H$
we obtain a model in which bosonic fluctuations of the total spin
interact with a band of spinless electrons.  In this limit it makes
sense first to diagonalise the Hund's rule coupling term in
the Hamiltonian and then to introduce the hopping of
electrons as a ``perturbation''.  We do this following the method 
introduced in
\cite{Shannon01b} by constructing new Fermi operators
$\{\f,\fd\} = 1$ and $\{\p,\pd\} = 1$ which create
eigenstates of the Hund's rule coupling term with eigenvalue $-J_HS/2$ and
$J_H(S+1)/2$, respectively.  The Hund's rule coupling then reads
\be
- \frac{J_{H}S}{2}
   \left[ \fd \f
   - \left(1 + \frac{1}{S} \right)
   \pd \p
   + \frac{\fd \f \pd  \p}{S}
   \right]
\en
where the sum over sites has been suppressed.
In the physically relevant limit limit $J_H \to \infty$, for less than half
filling, we can remove $\p$ Fermions from the problem entirely,
and rewrite the kinetic energy term in Eqn. \ref{eqn:DE} entirely in
terms of a band of spinless ($\f$) electrons interacting with
fluctuations of the total spin parameterised by the Bose operators
$[\at,\atd] = 1$.

\begin{figure}[tb]
\begin{center}
\leavevmode
\epsfxsize \columnwidth
\epsffile{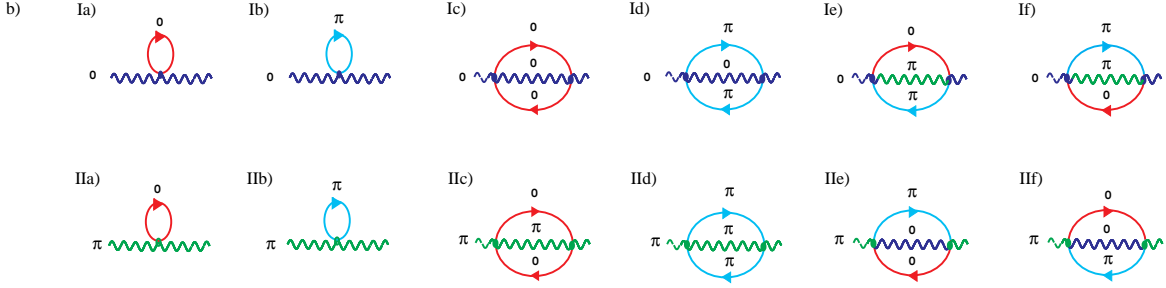}
\caption{
Leading self energy corrections for spin waves due to interaction
with electrons in limit $J_h/t \to \infty$.  Diagrams
Ia--f) show contributions for acoustic ($\at_{q0}$)
and diagrams IIa--f) optical ($\at_{q\pi}$ spin wave modes.
}
\label{fig:diagrams}
\end{center}
\end{figure}

To accomplish this transcription of the Hamiltonian it is sufficient
to know a few of the leading terms of the inverse transformation between
$c_{\uparrow}$
and $c_{\downarrow}$ electron operators, and the new $\f$ operators creating
eigenstates of the Hund's rule coupling term
\be
c_{\uparrow} &=& f\left(1 - \frac { \atd   \at}{4S}\right) +\ldots\\
c_{\downarrow} &=& \frac {f  \atd}{\sqrt{2S}}
   \left( 1 - \frac{1}{2 S}\right) +\ldots
\en
To prove this result, and to derive the full transformation
between ``laboratory frame'' $c_{\uparrow}$ and $c_{\downarrow}$ electron
operators and the ``local frame'' $\f$ and $\p$ operators,
together with the appropriate algebra for the spin boson $\at$
is an involved task.
We will not discuss the transformation in detail here
(see \cite{Shannon01a}), but note that
all the necessary canonical commutation and anticommutation relations,
\eg $[\f, \at] = 0$, \etc, are obeyed.

Up to a constant the transformed DE Hamiltonian reads
\be
\label{eqn:DEnew}
{\mc H} &=& {\mc H}_0 + {\mc V}_2 + {\mc O}(1/S^3)
\en
where the kinetic energy term for $\f$ electrons is
\be
{\mc H}_0 &=& - t \sum_{\langle ij\rangle \lambda}
      \fd_{i\lambda} \f_{j\lambda}
   - t_{\perp} \sum_{i} \left\{ \fd_{i 1} \f_{i 2}
      + \hc \right\}
\en

At this level the spin excitations $\at$ are dispersion-less.
Spin wave dispersion first enters into the problem through interaction at
${\mc O}(1/S)$ through the interaction term
\be
{\mc V}_{2} &=&
   - \frac{t}{4S}\sum_{\langle ij\rangle \lambda}
   \fd_{i\lambda} f_{j\lambda}
   \left[\left(
         \atd_{j\lambda} \at_{j\lambda} +
         \atd_{i\lambda} \at_{i\lambda}\right)
         \left(1-\frac {3}{8 S}
         \right)
         - 2 \atd_{i\lambda} \at_{j\lambda}
         \left( 1 - \frac{1}{2 S}\right)
   \right]\no\\
   &&\qquad
   - \frac{t_{\perp}}{4S}\sum_{i}
   \left\{
   \fd_{i1} f_{i2}
   \left[\left(
         \atd_{i1} \at_{i1} +
         \atd_{i2} \at_{i2}\right)
         \left(1-\frac {3}{8 S}
         \right)
         - 2 \atd_{i1} \at_{i2}
         \left( 1 - \frac{1}{2S}\right)
   \right] + \hc
   \right\}
\en
where we have neglected a further four boson vertex at ${\mc
O}(1/S^2)$ which is irrelevant at zero temperature.

By Fourier transformation we obtain the following Hamiltonian
which describes a band of spinless electrons interacting with
(initially dispersion-less) bosonic spin--wave excitations.
\be
\label{eqn:hamiltonian}
{\mc H} &=& {\mc H}_0 + {\mc V}_{2} + {\mc O}(1/S^3)\no\\
{\mc H}_0 &=& \sum_{k}
   \left(\epsilon_k - t_{\perp}\right)
   \fd_{k0} \f_{k0}
   + \left(\epsilon_k + t_{\perp}\right)
   \fd_{k\pi} \f_{k\pi}\\
{\mc V}_{2}
   &=& \frac{1}{N} \sum_{k_{1} \cdots k_{4}}
         \sum_{\eta \eta^{\prime} \nu \nu^{'}=\{0,\pi\}}
         {\mc V}_{2\eta^{'}4\nu^{'}}^{1\eta 3\nu}
         \fd_{1\eta} f_{2 \eta^{\prime}}
         \atd_{3 \nu} \at_{4\nu}
         \delta_{1+3-2-4}
         \delta_{\eta+\nu-\eta^{'}-\nu^{'}}\no
\en
where we consider symmetric and antisymmetric combinations of electron
operators (binding and antibinding bands), and of spin operators
(acoustic and optical spin waves), for the two planes.
For the simple nearest neighbour tight--binding model
Eqn. \ref{eqn:DE} the in--plane electronic dispersion is given by
$\epsilon_k$ = - zt$\frac{1}{2}(\cos k_x+\cos k_y)$ in units where
the distance between Mn atoms $a=1$.
The scale of interaction between electrons and spin waves
${\mc V}_{2}$ is determined entirely by electronic energies,
but is one order in $S$ down on the kinetic energy term ${\mc H}_0$.
There are a total of eight physically distinct vertices (decay
channels) for interaction between electrons and spin excitations.
The convention for labelling these vertices is shown in Fig.~\ref{fig:vertex} 
and their algebraic expressions are given in
Eqn.~\ref{eqn:VE1} and \ref{eqn:VE2}. 
The spin wave dispersion is now determined
by the leading order self energy up to 1/S$^2$ shown in
Fig. \ref{fig:diagrams}. We first discuss the results within the
usual semiclassical (1/S) approximation.

\section{Theoretical Predictions}

\label{theory}

In a cubic system, at a semi--classical level of approximation,
Zener's DE mechanism leads to a FM effective nearest neighbour
Heisenberg exchange interaction between neighbouring Mn spins,
with a spin wave dispersion
\be
\label{eqn:dispersioncubic}
\omega_q = zJ^{DE}S
   [1 - \gamma_q]
\en
where the size of the effective exchange interaction is set by electron
energies \cite{Furukawa96}
\be
\label{eqn:JDE}
J^{DE} = \frac{1}{2S^2}\frac{t}{N} \sum_{k} \gamma_k n(k)
\en
Here $\gamma_q = \frac{1}{3}(\cos q_x +\cos q_y +\cos q_z)$ is the
structure factor for a 3D cubic lattice and
$n(k)$ is the occupation of the electronic state with
momentum $k$, and J$^{DE}$ is proportional to the
expectation value of the kinetic energy operator
per Mn--Mn bond, relative to the center of the band.
Spin waves are exact eigenstates of a Heisenberg FM,
and therefore undamped.
This simple mapping between DE and Heisenberg models breaks down,
however, when quantum effects are included \cite{Shannon01b}.

\begin{figure}[tb]
\begin{center}
\leavevmode
\epsfysize = 75.0mm
\epsffile{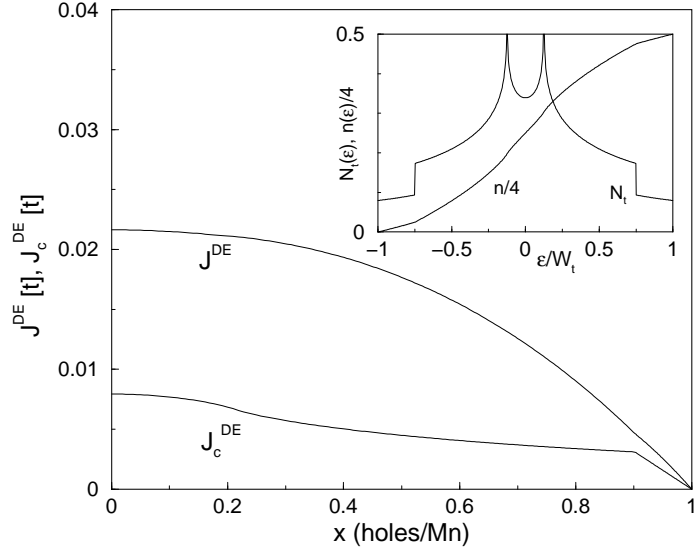}
\caption{
Inset shows total DOS N$_t$($\epsilon$)= N$_0$($\epsilon$)+
N$_\pi$($\epsilon$) and total electron number n for the 2D bands
$\epsilon_0(k)$ and $\epsilon_\pi(k)$ W$_t$=W+t$_\perp$ is half of the
overall bandwidth. Main figure shows variation of J$^{DE}$ and
J$_c^{DE}\equiv$ J$_\perp^{DE}$ with hole doping under the
assumption that t$_\perp$/t=0.57 (fixed for x=0.4) is {\em independent} of x.
The physical FM regime is restricted to 0.3$<x<$0.4.
}
\label{fig:DOS}
\end{center}
\end{figure}

The situation in a bilayer system is complicated in that
there are both optical and acoustic
branches of spin wave excitations, but the mapping
onto an effective Heisenberg model is once again possible at
a semi--classical level.
Evaluating the effect of interaction between electrons and
spin waves described by Eqn. \ref{eqn:hamiltonian}
to ${\mc O}(1/S)$, and now including the effect
of super--exchange terms, we obtain a spectrum :
\be
\label{eqn:dispersionbilayer}
\omega^0_{q} &=& z(J^{DE} + J^{EX})S[1 - \gamma_q] \no\\
\omega^{\pi}_{q} &=& z(J^{DE} + J^{EX})S[1 - \gamma_q]
   + 2(J^{DE}_{\perp}+J^{EX}_{\perp})S
\en
where in 2D $\gamma_q = \frac{1}{2}(\cos q_x+\cos q_y)$,
$\omega^0_{q}$ is the dispersion of the acoustic
and $\omega^{\pi}_{q}$ the dispersion of the optical
spin wave branch.
The size of the DE in-plane contribution to the effective exchange
integral is once again set by the expectation value of the kinetic
energy on a single bond, and the DE between the two planes is
determined by the occupation difference of binding and antibinding bands:
\be
\label{eqn:biex}
J^{DE} = \frac{1}{2S^2}\frac{t}{2N} \sum_{k} \gamma(k)
         \left[n_{0}(k)+n_{\pi}(k)\right]\no\\
J^{DE}_{\perp} = \frac{1}{2S^2}\frac{t_{\perp}}{2N} \sum_{k}
                 \left[n_{0}(k)-n_{\pi}(k)\right]
\en
Here we used the occupation numbers $n_0(k)$ =$\langle
f^\dagger_{k0}f_{k0}\rangle$ and $n_{\pi}(k)$ =$\langle
f^\dagger_{k\pi}f_{k\pi}\rangle$ of the binding and antibinding
electron bands $\epsilon_0(k)$= -t$_\perp$+$\epsilon(k)$,
$\epsilon_{\pi}(k)$= t$_\perp$+$\epsilon(k)$ respectively. Our result at
this order agrees perfectly with earlier calculations of the spin wave
spectrum in a bilayer \cite{Chatterji99a,Chatterji99b}. The effective exchange
constants in
Eqn. \ref{eqn:biex} can be evaluated as function of the doping x which
gives the number of holes per Mn-site or the total number of e$_g$ electrons
per Mn site n= n$_0$+n$_\pi$= 1-x that occupy the 0,$\pi$- bands.
By using the DOS functions N$_{0,\pi}(\epsilon$)=N($\epsilon\pm
t_\perp$), the electron number n$_{0,\pi}$ and the average band energy
$\epsilon_{0,\pi}$ of the 2D binding and antibinding bands respectively may
be expressed as
\be
\label{eqn:dens}
n_{0,\pi}&=&\int_{-W}^{\epsilon_F\pm t_\perp}N(\epsilon)d\epsilon\no\\
\epsilon_{0,\pi}&=&\int_{-W}^{\epsilon_F\pm t_\perp}
N(\epsilon)\epsilon d\epsilon\\
N(\epsilon)&=&\frac{2}{\pi^2}\frac{1}{W}
K([1-(\frac{\epsilon}{W})^2]^\frac{1}{2})\no
\en
Here W=zt and 2W is the band width of of each of the 2D bands
$\epsilon_{0,\pi}(k)$ and $\epsilon_F$ is the Fermi level. Furthermore
K($\xi$) is the complete elliptic integral of the first kind. The
total DOS N$_t$=N$_0$+N$_\pi$ and the total number of
electrons n as a function of the Fermi level is shown in the inset of
Fig. \ref{fig:DOS}. The spikes in the DOS are logarithmic
singularities of each of the 2D bands at its band center
($\pm$t$_\perp$). We then obtain for the effective DE exchange constants
after Eqn. \ref{eqn:biex}:

\begin{figure}[tb]
\begin{center}
\leavevmode
\epsfysize = 75.0mm
\epsffile{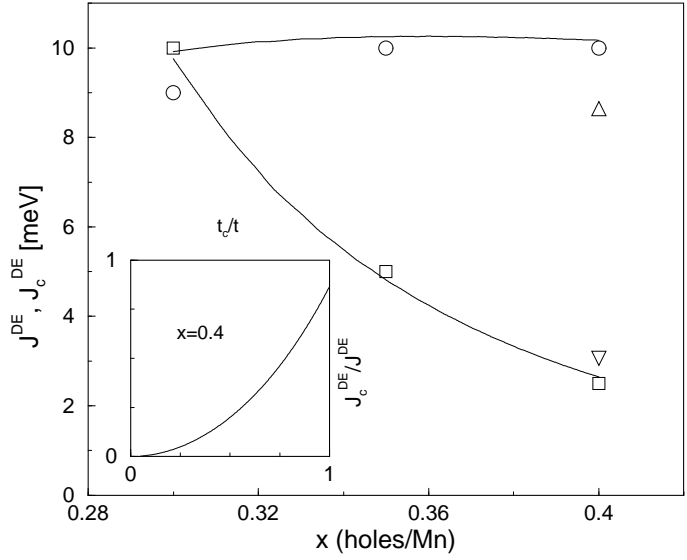}
\caption{
Doping dependence of DE exchange constants. The strong reduction of
J$_c^{DE}\equiv$J$_\perp^{DE}$ with increasing x is due to the strong
reduction of t$_c\equiv$t$_\perp$ due to the JT effect of MnO units as
described by Eqn. \ref{eqn:gruen} with $\Omega_\perp\eta_\perp$=-2.
Experimental data from \cite{Chatterji99b} (circles),
\cite{Hirota01} (squares) and \cite{Perring01} (triangles). 
In the latter case a somewhat smaller SJ$^{DE}$=8.6 meV was
obtained by fitting the spin wave stiffness constant in contrast to the
value of 10 meV obtained by fitting to the whole spin wave bandwidth
W$_{sw}^{[100]}$=zSJ$^{DE}$.
The inset shows the dependence of DE exchange anisotropy on the
hopping anisotropy for a fixed doping x=0.4.
}
\label{fig:doublex}
\end{center}
\end{figure}

\be
\label{eqn:double}
J^{DE}&=&-\frac{1}{2S^2}\frac{1}{2z}(\epsilon_0+\epsilon_\pi)\no\\
J^{DE}_{\perp}&=&\frac{1}{2S^2}\frac{t_\perp}{2}(n_0-n_\pi)
\en

The DE anisotropy ratio $J^{DE}_{\perp}/J^{DE}$ is is equal to the
ratio $\omega_0^\pi$/W$_{sw}^{[110]}$
of optical spin wave gap $\omega_0^\pi$=2SJ$_\perp^{DE}$ to the
acoustical (or optical) spin wave band width W$_{sw}^{[110]}$=
2zSJ$^{DE}$ along [110] direction  and it is given by
\be
\label{eqn:anisotropy}
\frac{J^{DE}_{\perp}}{J^{DE}}&=&-(\frac{t_{\perp}}{t})
\frac{W(n_0-n_\pi)}{\epsilon_0+\epsilon_\pi}
\en

For t$_{\perp}\ll t$ the occupation number difference increases linearly 
in t$_{\perp}$, and using Eqn. \ref{eqn:anisotropy} we find
J$^{DE}_{\perp}$/J$^{DE}\sim(t_{\perp}/t)^2$, as shown in the
inset of Fig. \ref{fig:doublex}.  Numerical values from
Eqs. \ref{eqn:double} are presented in Fig. \ref{fig:DOS} as function
of the doping x=1-n. It shows the variation for the model DE for fixed
t$_\perp$ in the whole range 0$\leq x\leq 1$ although it must be kept
in mind that the physical region for the FM phase of
La$_{2-2x}$Sr$_{1+2x}$Mn$_2$O$_7$ is much smaller, according
to \cite{Hirota01} it exists for 0.3$\leq x\leq$ 0.4.
From a comparison of the experimental values of the
optical spin wave gap and the spin wave band width at x=0.4 with
Eqn. \ref{eqn:double} and with the insert in Fig. \ref{fig:doublex} we
can obtain estimates for the underlying microscopic model parameters
within the classical approximation, namely t$_\perp$/t$\simeq$ 0.57
corresponding to the experimental J$^{DE}_{\perp}$/J$^{DE}\simeq$ 0.30
at low temperature and t$\simeq$ 0.175 eV (t$_\perp$= 0.1 eV) as
obtained from the experimental value SJ$^{DE}$= 10 meV (from
W$_{sw}^{[100]}$= zSJ$^{DE}$= 40 meV \cite{Chatterji01a}) by using
Eqn. \ref{eqn:double}.
According to Fig. \ref{fig:DOS} J$^{DE}$(x) and
J$^{DE}_\perp$(x) should not change dramatically with the hole doping
in the FM regime 0.3 $\leq x\leq$ 0.4, namely at most 6\% and 15\%
respectively. However this refers to the artificial situation where
t$_\perp$ does not depend on the doping. From the
experimental investigation of optical spin wave gap
and dispersion for various dopings x=0.30, x=0.35 and x=0.40
\cite{Perring01} it is known that indeed J$^{DE}$ shows no change in
this region, however J$^{DE}_\perp$(x) strongly increases by a
factor of four when the doping is reduced from x=0.4 to
x=0.3. 

The origin of this pronounced doping dependence of interlayer DE is
connected to the large Jahn-Teller(JT) distortion
observed\cite{Kubota00} in the bilayer manganites. This distortion is
defined as $\Delta_{JT}$=apical Mn-O bond length/equatorial bond
length. Decreasing the doping leads to an increase of
$\Delta_{JT}$. The driving mechanism for this JT distortion is an
increasing admixture of d$_{3z^2-r^2}$ states into the conduction band
states which naturally leads to an increase of t$_\perp$ with reduced
doping, which in turn strongly increases the interlayer
J$^{DE}_\perp$(x) as shown in the inset of
Fig. \ref{fig:doublex}. For the limited FM doping range one may
describe this dependence by introducing dimensionless Gr\"uneisenparameters
$\eta_\perp$=-($\partial\ln D/\partial\ln x$) and
$\Omega_\perp$=-($\partial\ln t_\perp/\partial\ln D$)
where D= distance between the layers of a single bilayer. They
describe the doping dependence of the JT- distortion and the
distortion dependence of the interlayer- hopping respectively. The
JT effect on the intra-layer hopping t is neglected since no
doping dependence of J$^{DE}$ is
observed. Assuming that $\eta_\perp$ and $\Omega_\perp$ are constants
in the range of x considered, this amounts to a doping dependence of
t$_\perp$ given by
\be
\label{eqn:gruen}
t_\perp(x)=t^0_\perp(\frac{x}{x_0})^{\Omega_\perp\eta_\perp}
\en
where e.g. x$_0$=0.4 and t$_\perp^0$=t$_\perp$(x$_0$).
According to the physical origin of the JT distortion mentioned above
one has to expect that $\Omega_\perp\eta_\perp <0$. Using the above
relation in Eqs. \ref{eqn:double} with $\Omega_\perp\eta_\perp$= -2
one obtains the doping dependence of
the exchange constants shown in Fig. \ref{fig:doublex} together with
the experimental values for various dopings. Using $\eta_\perp\simeq$ 0.037
from the JT- distortions given in \cite{Kubota00} one then obtains from the
above relation $\Omega_\perp\simeq$ -54. This large negative
Gr\"uneisenparameter characterises a strong dependence of the
effective interlayer hopping t$_\perp$ on the layer spacing D within a
bilayer. The JT-- distortion increases
with temperature for constant doping implying an increase of t$_\perp$
and hence J$^{DE}_\perp$. The DE therefore becomes more isotropic at
higher temperature. This has indeed been observed for x=0.4 in diffuse neutron
scattering where J$^{DE}_\perp$/J$^{DE}\simeq$ 0.5  has been found at
room temperature\cite{Chatterji01b} compared to
J$^{DE}_\perp$/J$^{DE}\simeq$ 0.3 from the low temperature spin wave
experiments discussed here \cite{Jackeli01}.

\begin{figure}[tb]
\begin{center}
\leavevmode
\epsfysize = 75.0mm
\epsffile{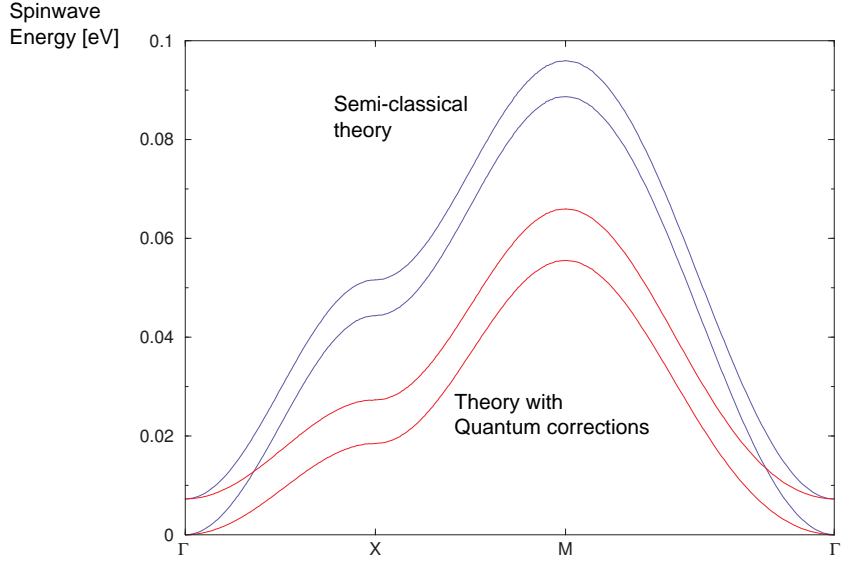}
\caption{Spectrum for optical and acoustic spin wave modes
throughout the Brillouin zone, calculated for doping $x=0.4$,
$t=0.175eV$, $t_{\perp}= 0.1eV$.   
Upper pair of lines ---
semi--classical spin wave dispersion at ${\mc O}(1/S)$;
lower pair of lines --- spin wave spectrum including quantum effects
at ${\mc O}(1/S^2)$.}
\label{fig:dispersion}
\end{center}
\end{figure}

\begin{figure}[tb]
\begin{center}
\leavevmode
\epsfysize = 75.0mm
\epsffile{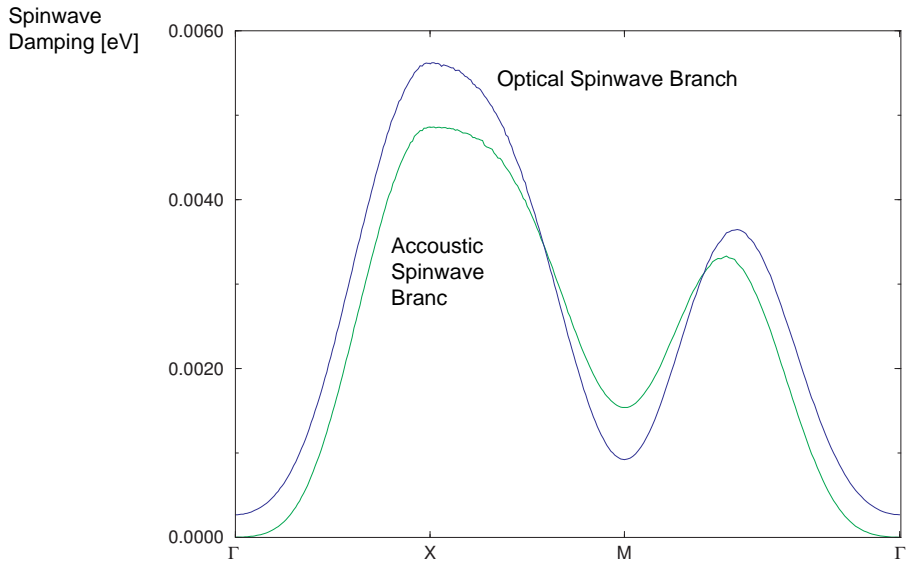}
\caption{Damping of acoustic and optical spin wave modes, throughout
the Brillouin Zone, calculated for doping $x=0.4$,
$t=0.175eV$, $t_{\perp}= 0.1eV$.}
\label{fig:damping}
\end{center}
\end{figure}

The results for the dispersion of acoustic and optic
spin wave modes for throughout the Brillouin zone, at a semi--classical
level and for the parameters given above are shown in 
Fig.~\ref{fig:dispersion}.  At zero temperature, at a 
semi--classical level, spin waves are
undamped, \ie states with a single spin wave excitation are good
eigenstates of the Hamiltonian, with no allowed decay processes.
The dispersion of spin waves is generated by their {\it elastic}
scattering by the average density of electrons.
At ${\mc O}(1/S^2)$ spin waves can decay through interaction
with electrons into lower energy spin excitations dressed with
particle hole pairs --- an inelastic process.  
This leads to a damping of spin waves,
and a corresponding shift in spin wave dispersion to lower
energy.  We can evaluate both effects starting from the Hamiltonian
Eqn.~\ref{eqn:hamiltonian}.  We consider

\be
\bar\omega^{\nu}_q &=& \omega^{\nu}_q + Re\{\Sigma^{\nu} (q,0)\}\no\\
\gamma^{\nu}_q &=& - Im\{\Sigma^{\nu} (q,\omega^{\nu}_q)\}
\en
where $\bar\omega^{\nu}_q$ is the net dispersion and
$\gamma^{\nu}_q$ the damping of the spin wave excitation,
and $\Sigma^{\nu}(q,\Omega)$ is the momentum and frequency dependent
self energy correction due to interaction of spin waves with electrons
at ${\mc O}(1/S^2)$.
The various contributions to the spin wave self energy are shown in
Fig.~\ref{fig:diagrams} at this order and given in Appendix~\ref{selfenergy}. 
The new physical process involved is the {\it inelastic} 
scattering of spin waves from fluctuations of charge density.
Results for spin wave dispersion, including leading quantum
corrections are shown in Fig.~\ref{fig:dispersion}, and values
for the damping of spin waves in Fig.~\ref{fig:damping}.

The immediate conclusion which we can draw from these
calculations is that quantum effects on spin waves
in a DE bilayer are very large.
The downward renormalisation of spin wave dispersion
at ${\mc O}(1/S^2)$ is a sizable fraction (about 30\%) of
the spin wave dispersion at ${\mc O}(1/S)$.
Similarly, the damping of spin waves is quite pronounced, being of the
scale 5--10\% of the spin wave dispersion, rising to a maximum
value $\sim 6$ meV at the zone corner.
Because of the large renormalisation of the spin wave
spectrum at ${\mc O}(1/S^2)$ it would be necessary to
reparameterise our model to fit experimental data
with the leading quantum effects included, by increasing
the sizes of the hopping integrals $t$ and $t_{\perp}$,
and including super-exchange interactions $J^{EX}$ and $J^{EX}_{\perp}$.
Any increase in the electron
bandwidth would give a proportionate increase in the
damping of spin waves.

Examining the quantum corrections in more detail, we find
that the spin wave dispersion has been modified so as
to give a relative softening of spin wave modes near the
zone center.
This can be understood loosely in terms of the
dynamical generation of an effective non nearest--neighbour
couplings between spins by processes ${\mc O}(1/S^2)$.
It is also interesting to note that the gap between
acoustic and optical modes is now momentum
dependent.

While these effects are of themselves interesting, they do not
offer any unambiguous signatures of quantum effects in the magnetism
of La$_{2-2x}$Sr$_{1+2x}$Mn$_2$O$_7$, as one could achieve
similar modifications of the spin wave dispersion simply by 
postulating additional exchange couplings between spins on an
{\it ad hoc} basis.  It is the damping of spin waves at zero 
temperature which sharply distinguishes a DE system from any
conceivable Heisenberg ferromagnet.   In a the Heisenberg 
ferromagnet, spin waves are undamped at zero temperature, 
and damping only becomes appreciable for temperatures 
large compared with the spin stiffness D. 

The damping which we predict for the DE model Eqn.~\ref{eqn:DE}
at zero temperature is large and highly momentum dependent.
The zone centre acoustic mode must remain undamped (in the absence
of any magnetic anisotropy) as it corresponds to the rotation of the 
total magnetization of the is the Goldstone mode of the system.
Accordingly, in the zone centre, we find that the damping of the 
acoustic mode vanishes as
\be
\Gamma^{AC}_q = \alpha^{AC} q^5
\en
The optical spin waves are not Goldstone modes, however, and
have a finite dispersion (and damping) in the zone centre.
We find that the latter behaves as 
\be
\Gamma^{OP}_q = \Gamma^{0P}_0  + \alpha^{OP} q^3
\en
The lower power law in $q$ here reflects the way in which the vertex
for spin wave scattering is cut off by the interlayer hopping $t_{\perp}$.

Away from the zone centre the spin wave damping exhibits stationary 
points at the symmetry points of the brillouin zone --- a maximum
for both acoustic and optic modes at $X$, and a minimum for 
both at $M$.  It is interesting to note that the higher energy optic
modes are not always more strongly damped than the acoustic modes,
and that the maximum damping does not occur for the highest spin wave 
energies, as one might expect.  Infact the momentum dependence of the 
damping of spin waves in DE systems varies strongly with doping,
being constrained by both the geometry of the Fermi surface and 
the complex momentum dependence of the spin wave scattering vertex.

\section{Comparison with experiment}
\label{experiment}

The spin wave dispersions in bilayer manganites have been
investigated by several groups
\cite{Hirota01,Chatterji99a,Chatterji99b,Chatterji01a,Fujioka99,Chaboussant00,Perring01},
and a consensus was reached that the data could not be 
explained using a nearest neighbour Heisenberg model dispersion
of the form Eqn.~\ref{eqn:dispersionbilayer}
\cite{Hirota01,Chatterji99b,Chatterji01a,Fujioka99}.
Departures from Heisenberg model behaviour also have been observed 
many cubic manganite systems, for example 
Pr$_{0.63}$Sr$_{0.37}$MnO$_3$ \cite{Hwang98}.
Typically, what has been seen in both cubic and bilayer systems
is a softening and broadening of the zone boundary spin waves.
The total spin wave bandwidth measured to the zone boundary
is much less than would be predicted on the basis of the spin 
stiffness $D$ measured in the zone center, and the zone boundary
spin wave modes are extremely broad in comparison with their
energy.
The theory of bilayer manganites presented here shows that the 
double exchange model can exhibit both of these effects,
when quantum corrections are included.
However, the minimal model Eqn.~\ref{eqn:DE},
as parameterized above, is not sufficient to obtain
a quantitative description of the experimental results.

With regard to the spin wave dispersion, the inclusion of the 
quantum corrections shown in Fig~\ref{fig:dispersion} cannot 
explain the measured departures from Eqn.~\ref{eqn:dispersionbilayer}.  
The dispersion measured by inelastic neutron scattering 
on La$_{1.2}$Sr$_{1.8}$Mn$_2$O$_7$ \cite{Chatterji01a},
for the acoustic mode in the zone center, has the form
\be
\omega^0_q = \Delta + D^0q^2
\en  
with $\Delta \ll D^0$, as would be expected for a FM with 
small magnetic anisotropy.  
However at larger momentum transfer the measured 
dispersion lies below the curve
\be
\label{eqn:experiment}  
\omega^0_q = zD^0[1-\gamma_q]
\en
(away from the zone center we can safely neglect $\Delta$).
The total acoustic spin wave bandwidth, as defined by the spin wave
energy at the zone corner, is about $15\%$ less than $2zD$.
If we compare with a suitably parameterized quantum theory for 
the DE model, we find that the predicted dispersion lies {\it above} 
the curve Eqn.~\ref{eqn:experiment}, and so the observed softening 
effect is absent --- infact quantum corrections have the wrong
``sign''.   This failure of the minimal model Eqn.~\ref{eqn:DE}
is neither very surprising nor very disappointing, given that we
have attempted to fit the spin wave dispersion of a complex
system with spin charge and lattice degrees of freedom 
throughout the Brillouin Zone, using only
two adjustable parameters.   However it is important to ask
which of the many simplifications made is to blame for 
this disagreement with experiment ? 
    
A better fit could probably be obtained
at a semi--classical level, by substituting a more realistic 
dispersion for the underlying electrons into the one loop diagrams
used to calculate the ${\mc O}(1/S)$ spin wave self energy.
In tight binding language, each hopping integral 
$t_{ij}$ has a corresponding DE coupling 
$J^{DE}_{ij}$ associated with it.  The inclusion 
of $t_{ij}$ beyond nearest neighbours to obtain a more
realistic electronic bandstructure therefore also modifies the 
form of dispersion of the classically equivalent effective 
Heisenberg model.
Attempts to calculate spin wave dispersion directly from 
electronic structure suggest that this effect is 
important, and leads to a softening of zone boundary modes, 
at least in cubic systems \cite{Solovyev99}.

At a quantum mechanical level, since interactions between
spin waves are mediated by density fluctuations of the
electron gas, it would be more realistic to use a 
screened form of the charge susceptibility in which long range 
interactions were suppressed.  We anticipate that this would also 
tend to enhance the softening of zone boundary modes.
The inclusion of leading quantum corrections in ${\mc O}(t/J_H)$,
likewise leads to a softening of zone boundary spin waves 
\cite{Perkins01,Shannon02}.

Each of these improvements to the model would involve
the introduction of new parameters, which would need to be checked
against electronic structure and other experiments.
Since the stated aim of this paper is to explore the minimal model 
Eqn.~\ref{eqn:DE}, we will not discuss such refinements 
further here.
A more interesting possibility to explain the difference 
between experiment and theory would be that spin waves are coupled 
to orbital and/or lattice modes.   We return to this below.

\begin{figure}[tb]
\begin{center}
\leavevmode
\epsfysize = 75.0mm
\epsffile{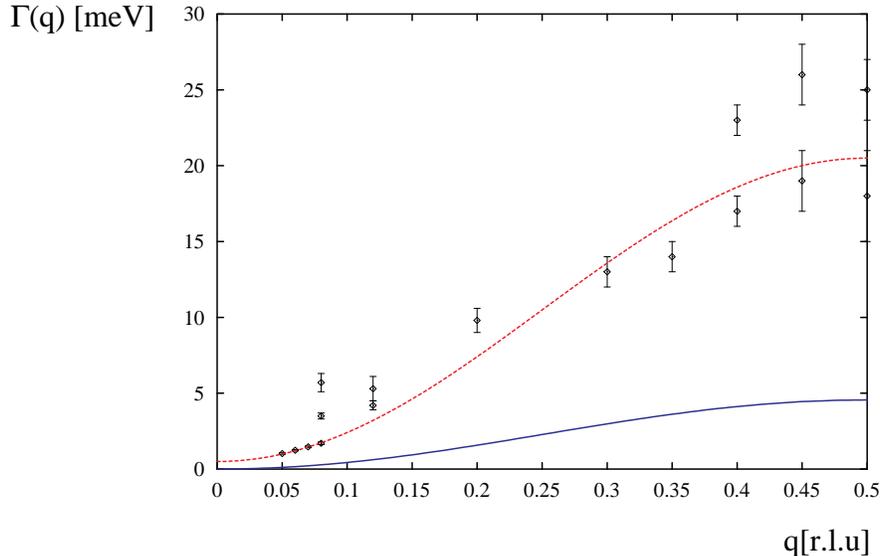}
\caption{Damping of acoustic spin wave modes on the line 
$\Gamma X$.  Points with error bars  --- experimental data
for La$_{1.2}$Sr$_{1.8}$Mn$_2$O$_7$ taken from \cite{Chatterji02};
lower solid line --- theoretical predicted damping for
minimal model with $x=0.4$, $t=0.175eV$, $t=0.1eV$.  The dashed
line is a guide to the eye.
}
\label{fig:tapan}
\end{center}
\end{figure}

The present theory is more
successful in explaining the damping at least in a qualitative way.
Fig.~\ref{fig:tapan} shows the experimentally observed 
widths of the energy scans as a function of momentum transfer $q$ from 
zone center to the zone boundary for
the acoustic spin excitations of bilayer manganite
La$_{1.2}$Sr$_{1.8}$Mn$_2$O$_7$,
along with that obtained from the present theory (continuous line). 
Data measured at different Neutron sources have been plotted together.

The damping which we calculate has a similar momentum dependence to that
observed, but is smaller by approximately a factor of four.   
Some part of the difference in absolute value between theory
and experiment can be explained by the fact that the 
parameters $t$ and $t_{\perp}$ were chosen so as to correctly 
reproduce the spin wave bandwidth at a semi--classical level,
and somewhat larger values must be used to fit the measured
dispersion once quantum effects are included, leading to some 
systematic variation in the values quoted and their associated 
errors.

Our conclusion is that the minimal DE model 
fails to explain the softening of zone boundary spin waves
in La$_{1.2}$Sr$_{1.8}$Mn$_2$O$_7$, but 
can explain about 30--40$\%$ of their width.
Of the refinements to the model discussed above, only the 
use of a screened charge susceptibility would 
affect the calculated damping of spin waves,
but we do not anticipate that this would lead to a marked
increase in their width.
We therefore conclude that spin waves
in the FM phase of this bilayer manganite are coupled 
dynamically to another mode, probably of orbital and/or lattice 
excitations.   
Such a couplings have been proposed in the context of the 
cubic manganites ---  for example to optical phonons 
\cite{Furukawa99} or through Jahn Teller active lattice 
modes to orbital fluctuations \cite{Khaliullin00}.

\section{Conclusions}
\label{conclusions}

We have constructed the simplest possible model for 
ferromagnetism in La$_{1.2}$Sr$_{1.8}$Mn$_2$O$_7$, 
based on Zener's double exchange mechanism within 
a one orbital picture for a single bilayer.
This model has two adjustable parameters, the 
intra-- and inter-plane hopping integrals $t$ and 
$t_{\perp}$.  At a semi--classical level it is 
equivalent to a Heisenberg model with intra-- and
interplane exchange integral $J^{DE}$ and $J^{DE}_{\perp}$.
The doping dependance of these parameters was
discussed, and the predictions of the 
effective Heisenberg model compared
with the results of inelastic Neutron scattering 
experiments.  

As the experiments show departures
from simple Heisenberg model behaviour in both 
the form of dispersion and the scale of damping
of the spin waves at low temperatures, we also calculated the 
leading quantum corrections to spin wave self
energies.  These arise because of the scattering of
spin waves from density fluctuations of the 
electron gas which are neglected in the semi--classical
approximation.
We find that the minimal
model considered cannot explain the softening of zone boundary 
spin wave modes, and
somewhat underestimates the damping of spin waves, even
when quantum corrections are included.
This suggests that spin waves are strongly coupled
to another low energy mode, presumably related to lattice
fluctuations, either by a direct coupling to phonons
or an indirect ``orbital fluctuation'' effect.

\section{Acknowledgments}

It is our pleasure to acknowledge helpful conversations with George Jackeli, 
Giniyat Khaliullin and Natasha Perkins.  This work was in part 
supported under the visitors program of MPI--PKS (N.\S.\ and F.\O.).

\newpage

\appendix

\section{Spinwave--electron interaction vertices and self energy
corrections at ${\mc O}(1/S^2)$}
\label{selfenergy}

First we give the interaction vertices V$_{2\eta' 4\nu'}^{1\eta 3\nu}$
in Fig. \ref{fig:vertex} and Eqn. \ref{eqn:hamiltonian}.
There are eight possible channels for electron--spin wave
interaction; these are labeled according to the convention
in Fig. \ref{fig:vertex}.
The coefficients of these vertices are given by
\be
\label{eqn:VE1}
{\mc V}_{02,04}^{01,03}
  &=& \frac{1}{4S} \left[ \frac{v_{24}^{13}}{2}
                        + \frac{t_{\perp}}{8S} \right]
\no\\
{\mc V}_{\pi2,\pi4}^{\pi1,\pi3}
   &=& \frac{1}{4S}
       \left[ \frac{v_{24}^{13}}{2}
            - 2t_{\perp}-\frac{t_{\perp}}{8S}\right]
\no\\
{\mc V}_{02,\pi4}^{01,\pi3}
   &=& \frac{1}{4S}
       \left[ \frac{v_{24}^{13}}{2}
            + 2t_{\perp}+\frac{t_{\perp}}{8S}\right]
\no\\
{\mc V}_{\pi2,04}^{\pi1,03}
   &=& \frac{1}{4S} \left[ \frac{v_{24}^{13}}{2}
                         - \frac{t_{\perp}}{8S}\right]
\no\\
{\mc V}_{\pi2,04}^{01,\pi3}
   &=& {\mc V}_{02,\pi4}^{\pi1,03}
       = \frac{1}{4S} \left[ \frac{v_{24}^{13}}{2}
                           + t_{\perp}\right]
\no\\
{\mc V}_{02,04}^{\pi1,\pi3}
   &=& {\mc V}_{\pi2,\pi4}^{01,03}
       = \frac{1}{4S} \left[ \frac{v_{24}^{13}}{2}
                            - t_{\perp}\right]
\en
Where the vertex depends on in--plane momenta only
through in--plane electronic dispersion
\be
\label{eqn:VE2}
   v^{13}_{24} &=&
   \left[\left( 1 - \frac{1}{2S}\right)
      \left (\epsilon_{1+3} + \epsilon_{2+4} \right)
      -  \left( 1 - \frac{3}{8S} \right)
      \left( \epsilon_1 + \epsilon_2 \right)
   \right]
\en
where $\epsilon_k = -zt\frac{1}{2}(\cos k_x + \cos k_y)$.
The fundamental energy scales in the DEFM are set by the kinetic energy
of the itinerant electrons, and so it is natural that the
electron spin wave scattering vertices are proportional to $t$/$t_{\perp}$.

Knowledge of the Hamiltonian Eqn. \ref{eqn:DEnew} is sufficient to develop a
zero temperature diagrammatic perturbation theory in $1/S$ for the spin wave
dispersion of the DE bilayer, up to ${\mc O}(1/S^2)$, and to calculate the
leading contributions to spin wave damping.
The relevant processes are shown in Fig. \ref{fig:diagrams}.
At ${\mc O}(1/S)$ only the single loop diagrams a) and b)
contribute.  These evaluate to give the Heisenberg--model
like result Eqn. \ref{eqn:dispersionbilayer} for the semi--classical
spin wave dispersion.

The one loop diagrams also contribute a constant
term and a further renormalisation of the classical dispersion at
 ${\mc O}(1/S^2)$, but all new quantum effects arise from the new
processes contributing to spin wave self energy at ${\mc O}(1/S^2)$,
the ``watermelon'' diagrams shown in Figure \ref{fig:diagrams} c)--f).
The self energy corrections for acoustic modes evaluate to give :
\be
\Sigma^{Ic} (\Omega,q)
   &=& \frac{1}{(4S)^2} \frac{1}{N^2}\sum_{kq^{'}}
     \left[
        \frac{zt}{2}
       \left( \gamma_{k}+\gamma_{k+q^{'}}
              - 2\gamma_{k+q} \right)
     \right]^2
     \frac{\theta\left(\xi_{k+q^{'}}^0\right)
           \theta\left(-\xi_k^0\right)
           }{
           \Omega-\omega_{q-q^{'}}^0-\xi_{k+q^{'}}^0+\xi_k^0+i\delta
           }
\\
\Sigma^{Id} (\Omega,q) &=&
   \frac{1}{(4S)^2}
   \frac{1}{N^2}\sum_{kq^{'}}
   \left[
      \frac{zt}{2}
      \left(
         \gamma_k+\gamma_{k+q^{'}}-2\gamma_{k+q}
      \right)
   \right]^2
   \frac{\theta\left(\xi_{k+q^{'}}^{\pi}\right)
         \theta\left(-\xi_{k}^{\pi}\right)
        }{
        \Omega-\omega_{q-q^{'}}^{0}-\xi_{k+q^{'}}^{\pi}+\xi_{k}^{\pi}+i\delta
        }
\\
\Sigma^{Ie} (\Omega,q)
   &=&\frac{1}{(4S)^2}
      \frac{1}{N^2}\sum_{kq^{'}}
      \left[\frac{zt}{2}
         \left(\gamma_{k}+\gamma_{k+q^{'}}
         - 2 \gamma_{k+q}\right)+t_{\perp}
      \right]^2
      \frac{
         \theta\left(\xi_{k+q^{'}}^0\right)
         \theta\left(-\xi_{k}^{\pi}\right)
      }{
        \Omega-\omega_{q-q^{'}}^{\pi}- \xi_{k+q^{'}}^0+\xi_{k}^{\pi}+i\delta}
\\
\Sigma^{If} (\Omega,q)
   &=&\frac{1}{(4S)^2}
      \frac{1}{N^2}\sum_{kq^{'}}
      \left[\frac{zt}{2}
         \left(\gamma_{k}+\gamma_{k+q^{'}}
         - 2 \gamma_{k+q}\right)-t_{\perp}
      \right]^2
      \frac{
         \theta\left(\xi_{k+q^{'}}^{\pi}\right)
         \theta\left(-\xi_{k}^{0}\right)
      }{
        \Omega - \omega_{q-q^{'}}^{\pi}-\xi_{k+q^{'}}^{\pi}+\xi_{k}^{0}+i\delta
      }
\en
where we have written out electron energies explicitly and
suppressed terms of ${\mc O}(1/S^3)$ in the vertex.
The corresponding processes for optical spin waves yield :
\be
\Sigma^{IIc} (\Omega,q) &=&
\frac{1}{(4S)^2}\frac{1}{N^2}\sum_{kq^{'}}\left
[\frac{zt}{2}\left(\gamma_{k}+\gamma_{k+q^{'}}-2\gamma_{k+q}\right)+2t_{\perp}
\right]^2
\frac{\theta\left(\xi_{k+q^{'}}^{0}\right)\theta\left(-\xi_{k}^{0}\right)}{
\Omega-\omega_{q-q^{'}}^{\pi}-\xi_{k+q^{'}}^{0}+\xi_{k}^{0}+i\delta}
\\
\Sigma^{IId} (\Omega,q) &=& \frac{1}{(4S)^2}
\frac{1}{N^2}\sum_{kq^{'}}
\left
[\frac{zt}{2}\left(\gamma_{k}+\gamma_{k+q^{'}}-2\gamma_{k+q}\right)-2t_{\perp}
\right]^2
\frac{\theta\left(\xi_{k+q^{'}}^{\pi}\right)\theta\left(-\xi_{k}^{\pi}\right)
}{
\Omega-\omega_{q-q^{'}}^{\pi}-\xi_{k+q^{'}}^{\pi}+\xi_{k}^{\pi}+i\delta}
\\
\Sigma^{IIe} (\Omega,q)
   &=&\frac{1}{(4S)^2}
      \frac{1}{N^2}\sum_{kq^{'}}
      \left[\frac{zt}{2}
         \left(\gamma_{k}+\gamma_{k+q^{'}}
         - 2 \gamma_{k+q}\right)+t_{\perp}
      \right]^2
      \frac{
         \theta\left(\xi_{k+q^{'}}^{\pi}\right)
         \theta\left(-\xi_{k}^0\right)
      }{
        \Omega- \omega_{q-q^{'}}^{0}-\xi_{k+q^{'}}^{\pi}+\xi_{k}^0+i\delta
      }
\\
\Sigma^{IIf} (\Omega,q)
   &=&\frac{1}{(4S)^2}
      \frac{1}{N^2}\sum_{kq^{'}}
      \left[\frac{zt}{2}
         \left(\gamma_{k}+\gamma_{k+q^{'}}
         - 2 \gamma_{k+q}\right)- t_{\perp}
      \right]^2
      \frac{
         \theta\left(\xi_{k+q^{'}}^0\right)
         \theta\left(-\xi_{k}^{\pi}\right)
      }{
        \Omega- \omega_{q-q^{'}}^{0}-\xi_{k+q^{'}}^0+\xi_{k}^{\pi}+i\delta
      }
\en
To ${\mc O}(1/S^2)$, we can neglect the frequency dependence of the
denominator in these expressions and evaluate the leading
quantum corrections to the dispersion of optical and acoustic
spin wave branches numerically by Monte Carlo integration.

If we restore the frequency dependence of the self energy
terms, we can also calculate the imaginary part of each.
We can use this to estimate the
spin wave damping on the mass shell, by setting the external
frequency equal to the semi--classical spin wave
dispersion at that wave number, \ie setting $\Omega = \omega^{0,\pi}_q$,
and eliminating all terms in the numerator of order spin wave
frequencies.
The contribution to damping from diagram IId) is, for example :
\be
\Gamma^{IId} (\omega^{\pi}_{q},q)
  &=&\frac{\pi}{(4S)^2}
   \frac{1}{N^2}\sum_{kq^{'}}
\left[zt_{\parallel}\left(\gamma_{k}-\gamma_{k+q}\right)
-2t_{\perp}\right]^2\nonumber\\
  &&\times\theta\left(\xi_{k+q^{'}}^{\pi}\right)\theta(-\xi_{k}^{\pi})
   \delta\left(\omega^{\pi}_{q}- \omega^0_{q-q^{'}} -
   \xi_{k+q^{'}}^{\pi}+\xi_{k}^{\pi}\right)
\en


\begin{references}
\bibitem{Jackeli01} 
Since completing this work we have 
become aware of a mean field theory calculation of 
the variation of $J^{DE}$ and $J_{\perp}^{DE}$ as function of 
doping for a two orbital model --- G.\ Jackeli and 
N.\ B.\ Perkins, cond--mat/0110185.
\end{references}
\end{document}